\documentclass[twocolumn]{aastex631}
\usepackage{stackengine}
\usepackage[encapsulated]{CJK}

\usepackage{float}
\usepackage{textcomp}

\begin{document}
\title{Remarkably Compact Quiescent Candidates at $3<z<5$ in JWST-CEERS}

\author[00000-0002-3722-322X]{Lillian Wright}
\affiliation{Department of Physics and Astronomy, Seoul National University, 1 Gwanak-ro, Gwanak-gu, Seoul 08826, Republic of Korea}
\affiliation{Department of Astronomy, University of Massachusetts Amherst, Amherst MA 01003, USA}

\author[0000-0001-7160-3632]{Katherine E. Whitaker}
\affiliation{Department of Astronomy, University of Massachusetts Amherst, Amherst MA 01003, USA}
\affiliation{Cosmic Dawn Center (DAWN), Denmark} 

\author[0000-0003-1614-196X]{John R. Weaver}
\affiliation{Department of Astronomy, University of Massachusetts Amherst, Amherst MA 01003, USA}

\author[0000-0002-7031-2865]{Sam E. Cutler}
\affiliation{Department of Astronomy, University of Massachusetts Amherst, Amherst MA 01003, USA}

\author[0000-0001-9269-5046]{Bingjie Wang (\begin{CJK*}{UTF8}{gbsn}王冰洁\ignorespacesafterend\end{CJK*})}
\affiliation{Department of Astronomy \& Astrophysics, The Pennsylvania State University, University Park, PA 16802, USA}
\affiliation{Institute for Computational \& Data Sciences, The Pennsylvania State University, University Park, PA 16802, USA}
\affiliation{Institute for Gravitation and the Cosmos, The Pennsylvania State University, University Park, PA 16802, USA}

\author[0000-0002-1482-5818]{Adam Carnall}
\affiliation{Institute for Astronomy, University of Edinburgh, Royal Observatory, Edinburgh, EH9 3HJ, UK}

\author[0000-0002-1714-1905]{Katherine A. Suess}
\altaffiliation{NHFP Hubble Fellow}
\affiliation{Kavli Institute for Particle Astrophysics and Cosmology and Department of Physics, Stanford University, Stanford, CA 94305, USA}
\author[0000-0001-5063-8254]{Rachel Bezanson}
\affiliation{Department of Physics \& Astronomy and PITT PACC, University of Pittsburgh, Pittsburgh, PA 15260, USA}
\author[0000-0002-7524-374X]{Erica Nelson}
\affiliation{Department for Astrophysical and Planetary Science, University of Colorado, Boulder, CO 80309, USA}
\author[0000-0001-8367-6265]{Tim B. Miller}
\affiliation{Center for Interdisciplinary Exploration and Research in Astrophysics (CIERA), Northwestern University, 1800 Sherman Ave, Evanston IL 60201, USA}

\author[0000-0002-9453-0381]{Kei Ito}
\altaffiliation{JSPS Research Fellow (PD)}
\affiliation{Department of Astronomy, School of Science, The University of Tokyo, 7-3-1, Hongo, Bunkyo-ku, Tokyo, 113-0033, Japan}

\author[0000-0001-6477-4011]{Francesco Valentino}
\affiliation{European Southern Observatory, Karl-Schwarzschild-Str. 2, 85748 Garching, Germany}
\affiliation{Cosmic Dawn Center (DAWN), Denmark} 

\begin{abstract}
In this letter, we measure the rest-frame optical and near-infrared sizes of ten quiescent candidates at $3<z<5$, first reported by \cite{carnall23}. We use \emph{James Webb Space Telescope} (JWST) Near-Infrared Camera (NIRCam) F277W and F444W imaging obtained through the public CEERS Early Release Science (ERS) program and \tt imcascade\normalfont, an astronomical fitting code that utilizes Multi-Gaussian Expansion, to carry out our size measurements. When compared to the extrapolation of rest-optical size-mass relations for quiescent galaxies at lower redshift, eight out of ten candidates in our sample (80\%) are on average more compact by $\sim$40\%. Seven out of ten candidates (70\%) exhibit rest-frame infrared sizes $\sim$10\% smaller than rest-frame optical sizes, indicative of negative color gradients. Two candidates (20\%) have rest-frame infrared sizes $\sim$1.4$\times$ larger than rest-frame optical sizes; one of these candidates exhibits signs of ongoing or residual star formation, suggesting this galaxy may not be fully quenched. The remaining candidate is unresolved in both filters, which may indicate an Active Galactic Nuclei (AGN). Strikingly, we observe three of the most massive galaxies in the sample (log(M$_{\star}$/M$_{\odot}$) = 10.74 - 10.95) are extremely compact, with effective radii ${\sim}$0.7 kpc. Our findings provide no indication that the size evolution relation flattens out, and may indicate that the size evolution of quiescent galaxies is steeper than previously anticipated beyond $z>3$.
\end{abstract}
\keywords{Quenched Galaxies (2016) --- Galaxy Formation (595) --- Galaxy Evolution (594)}
\section{Introduction} \label{sec:intro}

\begin{deluxetable*}{lccccccccc}
\tabletypesize{\scriptsize}
\tablecaption{Redshift, Stellar Mass and sSFR Estimate Comparison
\label{tab:comp}}
\tablehead{
\colhead{ID} & \colhead{ID} & \colhead{RA} & \colhead{DEC} & \colhead{\emph{z}$_{50}$}  & \colhead{\emph{z}$_{50}$}  & \colhead{log(M$_{\star}$/M$_{\odot}$)$_{50}$}  & \colhead{log(M$_{\star}$/M$_{\odot}$)$_{50}$}  & \colhead{log(sSFR/yr)$_{84}$}  & \colhead{log(sSFR/yr)$_{84}$} \\ 
\colhead{[Carnall+23]} & \colhead{[This Work]} & \colhead{} & \colhead{} & \colhead{[Carnall+23]}  & \colhead{[This Work]}  & \colhead{[Carnall+23]}  & \colhead{[This Work]}  & \colhead{[Carnall+23]}  & \colhead{[This Work]}}
\startdata
101962 & 68124 & 215.03906 & 53.00278 & 4.39$^{+0.11}_{-0.14}$ & 4.04$^{+0.18}_{-0.16}$ & 10.63$^{+0.06}_{-0.04}$ & 10.56$^{+0.09}_{-0.14}$ & -14.14 & -9.75$_{-1.91}$\\
97581 & 82632 & 214.98181 & 52.99124 & 3.46$^{+0.10}_{-0.09}$ & 3.42$^{+0.13}_{-0.09}$ & 10.81$^{+0.06}_{-0.03}$ & 10.84$^{+0.03}_{-0.03}$ & -13.14 & -11.71$_{-2.22}$\\
75768 & 82419 & 214.90485 & 52.93535 & 3.31$^{+0.10}_{-0.07}$ & 3.21$^{+0.06}_{-0.05}$ & 10.48$^{+0.07}_{-0.04}$ & 10.54$^{+0.02}_{-0.02}$ & -11.95 & -10.93$_{-1.08}$\\
36262 & 47006 & 214.89561 & 52.85650 & 3.26$^{+0.09}_{-0.10}$ & 3.20$^{+0.06}_{-0.05}$ & 11.06$^{+0.12}_{-0.07}$ & 10.83$^{+0.04}_{-0.07}$ & -14.66 & -9.33$_{-1.75}$\\
29497 & 89019 & 214.76063 & 52.84531 & 3.25$^{+0.08}_{-0.08}$ & 3.16$^{+0.05}_{-0.06}$ & 11.34$^{+0.06}_{-0.06}$ & 11.27$^{+0.03}_{-0.04}$ & -11.32 & -10.27$_{-1.55}$\\
92564 & 85652 & 214.95789 & 52.98030 & 3.47$^{+0.11}_{-0.10}$ & 3.51$^{+0.11}_{-0.10}$ & 10.49$^{+0.08}_{-0.04}$ & 10.45$^{+0.09}_{-0.06}$ & -11.49 & -10.08$_{-2.80}$\\
52175 & 71124 & 214.86605 & 52.88426 & 3.44$^{+0.14}_{-0.08}$ & 3.61$^{+0.09}_{-0.13}$ & 10.87$^{+0.05}_{-0.02}$ & 10.95$^{+0.03}_{-0.03}$ & -14.67 & -10.96$_{-1.52}$\\
42128 & 67843 & 214.85057 & 52.86603 & 4.19$^{+0.12}_{-0.12}$ & 3.59$^{+0.12}_{-0.11}$ & 11.06$^{+0.07}_{-0.05}$ & 10.23$^{+0.21}_{-0.23}$ & -14.56 & -7.92$_{-0.92}$\\
17318 & 65713 & 214.80817 & 52.83221 & 4.50$^{+0.13}_{-0.10}$ & 4.46$^{+0.09}_{-0.09}$ & 10.13$^{+0.04}_{-0.04}$ & 10.33$^{+0.04}_{-0.04}$ & -13.69 & -10.34$_{-0.56}$\\
8888 & 73305 & 214.76723 & 52.81771 & 3.49$^{+0.17}_{-0.12}$ & 3.47$^{+0.12}_{-0.08}$ & 10.54$^{+0.12}_{-0.08}$ & 10.53$^{+0.04}_{-0.07}$ & -12.83 & -9.69$_{-1.89}$\\
\enddata
\tablecomments{Median redshift, stellar mass and sSFR estimates for all ten candidates in the sample. This work adopts the Prospector-$\beta$ model \citep{Wang23}, whereas \citet{carnall23} utilizes \tt BAGPIPES\normalfont\ \citep{bagpipes}. 1${\sigma}$ errors are shown for redshift and stellar mass estimates. 1${\sigma}$ upper-bound estimates are shown for sSFR with 1${\sigma}$ lower-bound errors.}
\end{deluxetable*}

The size evolution of quiescent galaxies over cosmic time has been extensively studied and well-established up to $z=3$ \citep[e.g.,][]{Trujillo_2004,vanDokkum_2008,Kriek+08,cutler22,nedkova21}, and more recently extended to $z>3$ \citep{ito23, ormerod2023, ji2024jades}. Robustly defining and understanding the size evolution of quiescent galaxies provides us with insight regarding which physical mechanisms primarily contribute to the shut-down of star formation in galaxies (also known as \emph{quenching}). Quiescent galaxies are, on average,  much smaller than star-forming galaxies up to log(M$_{\star}$/M$_{\odot}$)$\sim$11 \citep{shen03, wel}. The exact physical processes that explain the compact morphology of quiescent galaxies at high redshift are still up for debate, but several theories could explain this correlation. One such theory is that compactness triggers quenching; when star-forming galaxies attain a certain central density or velocity dispersion threshold, a star-formation shutdown is triggered \citep{Franx_2008, Bell_2012, Omand_14, Tei_16, Whitaker_2017}. Taking measured size-mass relations at face value, one can then reverse engineer how quenching correlates with the suppression of star formation, black hole growth, and the role of halo concentration (e.g., \cite{chen+21}). The age of the universe at the redshift range we examine in our quiescent sample ($3<z<5$) is $\sim1-2$ Gyr, indicating that galaxies have already quenched in this early universe. Therefore, studying the morphology of these galaxies may illuminate the relationship between compact size and early quenching in quiescent galaxies. 

The evolving sizes of quiescent galaxies at $z>3$ reflect both the average evolution of the structures of the growing population and the intrinsic growth of individual quiescent galaxies. Quiescent galaxies experience a more rapid size evolution than star-forming galaxies, more than doubling in size from ‘cosmic noon’ ($z\sim1-2$) to the present day \citep{daddi05,vD08,damjanov09,mclure13,wel, hamadouche22}. Since these galaxies are no longer forming new stars or experiencing in situ growth, additional evolution - beyond what is expected to be added by late additions to the population - from more extended star-forming galaxies must be explained. A leading theory suggests that they grow in size mainly through the accretion of satellite galaxies or minor mergers \citep{Bezanson09, suess19}. In this model, quiescent galaxies follow an `inside-out' growth model that should result in an older (redder) stellar population in the central region and a younger (bluer) stellar population on the outskirts of the galaxy \citep{Bezanson09,Naab09,Hopkins09,vandeSande13,suess19}. This is known as a negative color gradient, which typically results in a noticeable size discrepancy between the light-weighted and true mass-weighted size measurements of the galaxy \citep{suess19}. This discrepancy appears most prominently in the rest-frame ultraviolet (UV) and is minimized in rest-frame near-infrared (NIR), which suggests that the latter is the most accurate tracer of stellar mass because it is less subject to changes in the mass-to-light ratio based on factors such as age or dust fraction \citep{Bell}. 

\begin{figure*}
\centering
\includegraphics[width=0.32\textwidth]{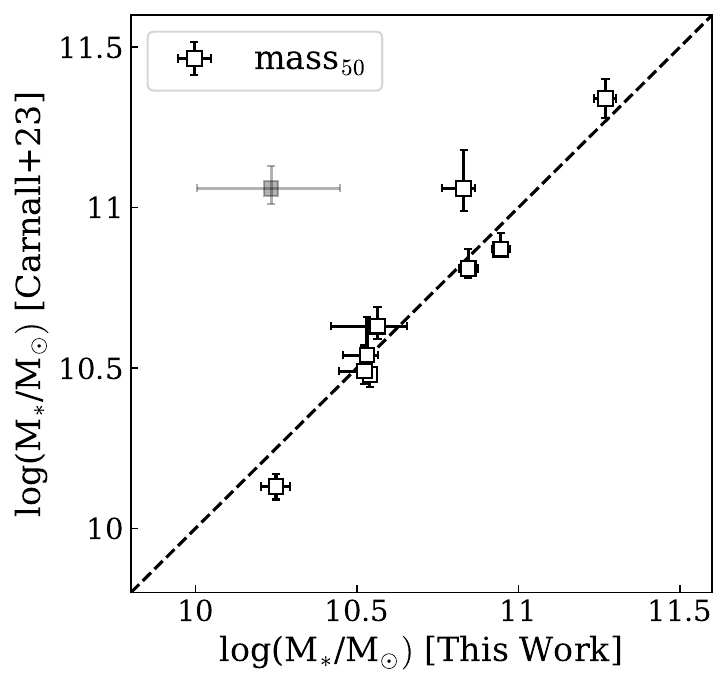}
\includegraphics[width=0.315\textwidth]{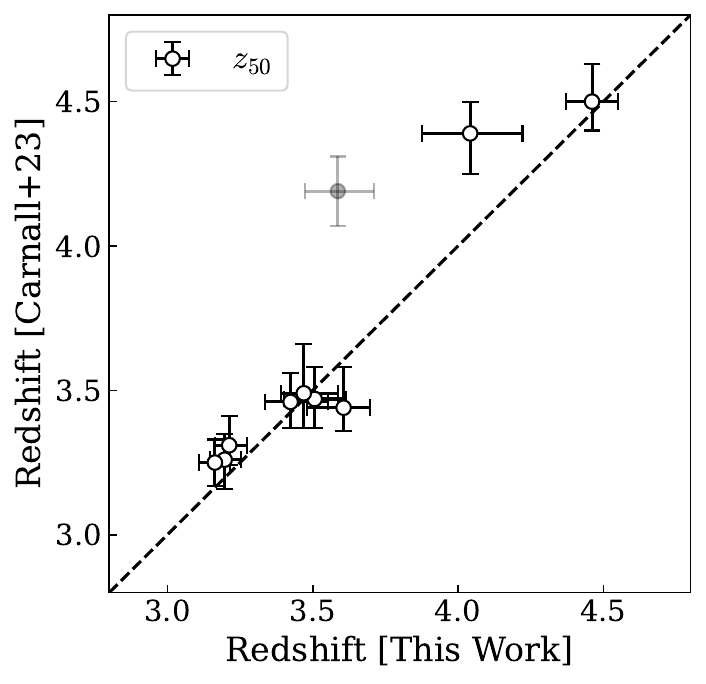}
\includegraphics[width=0.32\textwidth]{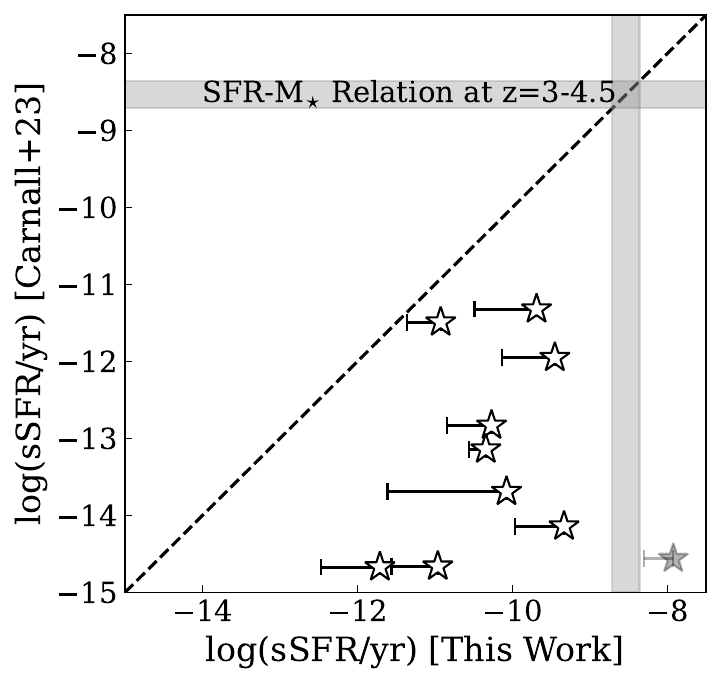}
\caption{The left and middle panels show the Prospector-$\beta$-produced median stellar mass and redshift estimates compared to the \tt BAGPIPES\normalfont-produced estimates reported by \cite{carnall23}. All error bars are 1${\sigma}$ and the dashed line shows a 1:1 ratio. The right panel shows the 1$\sigma$ upper-bound sSFR estimates produced by Prospector-$\beta$ and \tt BAGPIPES \normalfont with 1$\sigma$ errors. The shaded region marks the SFR-M$_{\star}$ relation for star-forming galaxies at $z=3.5-4$ from \cite{Speagle_2020}. The transparent point represents one candidate in the sample that exhibits signs of ongoing star formation, and thus may not be fully quenched.}
\label{fig:mass}
\end{figure*}

The longest wavelength filter on the \emph{Hubble Space Telescope} (HST), F160W (${\lambda}_\mathrm{obs}$ = 1.6$\mathrm{\mu}$m), probes the rest-frame optical light of galaxies at $z\sim$2 but migrates to rest-frame UV at higher redshifts. Therefore, with deep, high-resolution imaging from the \emph{James Webb Space Telescope} (JWST) beyond 2${\mu}$m, we are now able to obtain accurate rest-frame optical and NIR sizes of quiescent galaxies beyond $z>3$ for the first time. Studying the morphology of quiescent galaxies before the main epoch of quenching ($z>3$) when the universe was only $1-2$ Gyr old \citep[e.g.,][]{Gould_2023, antwidanso23, long23}, affords us the best opportunity to study the size evolution of quiescent galaxies and simultaneously observe the beginning of the quenching process before the effects of mergers dilute observational signatures \citep{ito23}.

In this letter, we examine the F277W (rest-frame optical, ${\lambda}_\mathrm{rest}$ = 0.46 - 0.69$\mu$m) and F444W (rest-frame NIR, ${\lambda}_\mathrm{rest}$ = 0.74 - 1.11$\mu$m) sizes of a sample of ten high-redshift ($3<z<5$) robust quiescent candidates first reported by \cite{carnall23}. Section~\ref{sec:data} presents the details of our data analysis, including the public release of our custom photometric catalog in the CEERS field for the sake of transparency and providing an independent resource to the community, as well as the details of size measurements and stellar population synthesis modeling. In Section~\ref{sec:res}, we extrapolate the expected size evolution for massive quiescent galaxies defined in \citet{wel} out to $z=5$ to compare with our \tt imcascade\normalfont-measured sizes. As an additional check on our results, we run our sample through \tt GALFIT \normalfont to ensure our size measurements are robust to any discrepancies in software and parametrization. We additionally incorporate the quiescent sample from \citet{suess22} to comprehensively examine the F150W, F277W, and F444W sizes of quiescent galaxies across $1<z<5$. The implications of our results are discussed in Section~\ref{sec:disc}. We assume a standard $\Lambda$CDM cosmology: ${\Omega}_{M}$=0.3, ${\Omega}_{\Lambda}$=0.7, and \emph{h}=0.7.

\section{Data \& Methodology} \label{sec:data}

\subsection{Sample Selection} 

Our targets consist of the ‘robust’ sub-sample defined in \cite{carnall23}, including ten quiescent candidates estimated within median photometric redshift $3<z_\mathrm{50}<5$ and stellar mass $10.13<\mathrm{log(M_{\star}/M_{\odot})}<11.34$. The ‘robust’ quiescent selection defined in \cite{carnall23} requires that 97.5\% of the specific Star Formation Rate (sSFR) posterior provided by SED-fitting code \tt BAGPIPES \normalfont \citep{bagpipes} satisfies the following criteria:

\begin{equation}
\textrm{sSFR}<\frac{0.2}{t_{\mathrm{obs}}},
\end{equation}

\noindent where t$_\mathrm{obs}$ is defined as the age of the universe at the redshift of the galaxy. For a more thorough overview of the quiescent selection criteria, refer to \cite{carnall23}. 

\subsection{Photometric Catalog}

The \citet{carnall23} study leverages the CEERS Early Release Science (ERS) program \citep{Finkelstein2017} over the All-wavelength Extended Groth Strip International Survey (AEGIS) field. For our purposes, we adopt publicly available JWST/NIRCam broadband imaging\footnote{\url{https://dawn-cph.github.io/dja/imaging/v7/}} covering a total area of $\sim$105 sq. arcmin produced with Grizli \citep{Brammer2023} (Please refer to \cite{Valentino_2023} for the latest image reduction documentation). We would like to note that these candidates were selected by \cite{carnall23} using only CEERS Epoch 1 data, which covers a total area of $\sim$30 sq. arcmin \citep{Bagley_2023}. Source detection and photometry are computed following \citet{weaver23}. In brief, sources are detected on a noise-equalized co-add of F277W, F356W, and F444W images. Photometry is extracted in 0.32\arcsec\, apertures from all available filters from HST (F435W, F606W, F814W, F105W, F125W, F140W, F160W) and JWST/NIRCam (F115W, F150W, F200W, F277W, F356W, F410M, F444W) over the AEGIS field, after psf-matching to F444W. Photometric errors are estimated by means of 10,000 aperture extractions on empty regions of the images (see \citealt{Whitaker2011}). To ensure robust stellar masses, the aperture photometry is corrected by means of Kron elliptical apertures measured on the F444W mosaic and further corrected by an additional $5-15$\% based on the light in the F444W beyond the circularized Kron apertures. An additional correction for the line-of-sight attenuation through the galaxy is then applied using the dust maps of \citet{Schlafly2011}. While this study is only focused on ten candidates within this field, the full catalog is available on \href{https://zenodo.org/doi/10.5281/zenodo.8400504}{Zenodo}. 

\subsection{Spectral Energy Distribution Fitting}

The SED-fitting analysis follows \citet{Wang2023:sps}, which is briefly reiterated here for completeness. All parameters, including redshift, are inferred jointly using the Prospector inference framework \citep{Johnson2021}, adopting the MIST stellar isochrones \citep{Choi2016,Dotter2016} and MILES stellar library \citep{Sanchez-Blazquez2006} from FSPS \citep{Conroy2010}. Star Formation History (SFH) is described non-parametrically via mass formed in seven logarithmically spaced time bins \citep{Leja2017}. Note that this differs from the double-power-law model used in \citet{carnall23}. A mass function prior, and a dynamic SFH($M, z$) prior are included to optimize the photometric inference for JWST observations (Prospector-$\beta$; \citealt{Wang23}). In total, the model consists of 18 free parameters, and sampling is performed using the dynamic nested sampler \texttt{dynesty} \citep{Speagle_2020}.

Our median redshift estimates fall within 1${\sigma}$ errors reported by \citet{carnall23} for eight out of ten candidates, with the exception of candidates 68124 and 67843\footnote{all candidate IDs listed in this letter correspond with our publicly available photometric catalog}. Similarly, our median stellar mass estimates fall within the combined 1${\sigma}$ errors provided by \tt BAGPIPES \normalfont and Prospector-${\beta}$ for seven out of ten candidates in the sample, with the exception of candidates 47006, 67843, and 65713. In general, we find that our  Prospector-$\beta$- redshift and stellar mass estimates show good agreement with those reported in \cite{carnall23}, with the exception of candidate 67843, which we predict to be at $z=3.59^{+0.12}_{-0.11}$ (as opposed to the \tt BAGPIPES \normalfont estimation of $z=4.19\pm0.12$). We additionally note that the median sSFR estimate of this candidate is log(SFR/M$_{\star}$)=-8.31$^{+0.38}_{-0.54}$ yr$^{-1}$, which is $\sim$0.1 dex above the log(SFR)-log(M$_{\star}$) relation at $z=3.59$ \citep[e.g.,][]{Speagle_2014} and thus indicative of possible residual or ongoing star formation.  Given this tentative evidence for ongoing star formation, we represent this particular target with increased transparency in all plots herein. All median redshift, median stellar mass, and 84th percentile sSFR estimates as well as the associated errors are listed in Table \ref{tab:comp}. These estimates and their errors are also illustrated in Figure \ref{fig:mass}.

\subsection{Morphological Measurements}

\begin{figure}
\centering
\includegraphics[width=0.5\textwidth]{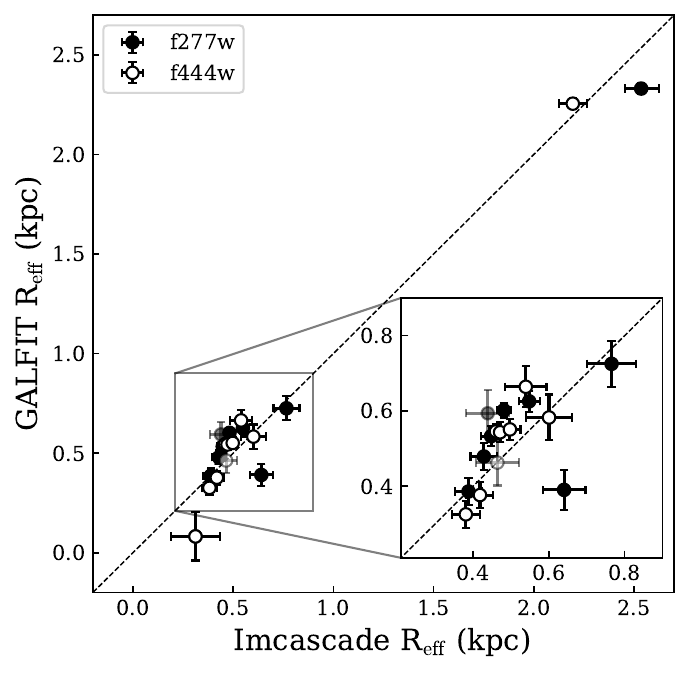}
\caption{Effective radii measured with \tt GALFIT \normalfont vs. \tt imcascade \normalfont from F277W (filled symbols) and F444W (open symbols) imaging. Errors are calculated by adding the software generated 1$\sigma$ errors in quadrature with a second term that scales with SNR in F160W. In general, we find broad agreement between the two independent size measurement methods, which further support out assertion that these quiescent candidates are extremely compact.}
\label{fig:galim}
\end{figure}

We fit for the effective radius in two of the available filters, F277W and F444W, selected to sample close to rest-frame 0.5$\mu$m and 1$\mu$m.  We use the astronomical fitting code \tt imcascade \normalfont \citep{miller}, which models the light profile of galaxies as a mixture of 2-D Gaussians in a Bayesian Framework. \tt imcascade \normalfont is optimal for studying the morphologies of faint, semi-resolved galaxies such as high-redshift quiescent galaxies \citep{miller}. First, we model the point spread function (PSF) as a Gaussian decomposition with four components to account for the complexity of the observed JWST PSF. As an additional check, we compare the curve-of-growth of our PSF model with the growth curves of both drizzled WebbPSFs and our observed CEERS PSFs \citep[e.g., see Appendix in ][]{weaver23} . We find that the width of the F444W and F277W PSFs at 50\% of the enclosed energy is consistent in all cases to within $\sim$1\% for the median redshift of our sample ($z\sim3.488$). We then use Source Extractor Python \tt sep \normalfont (\citealt{barbary16}, \citealt{sex}) to detect and mask sources in the image, as well as \tt imcascade's \normalfont ‘expand mask’ feature to convolve the mask with a Gaussian and shrink or expand the mask, ensuring contamination from neighboring objects is minimized. We use a cutout size of $\sim$40$\times$ the estimated Kron radius for each galaxy to properly model the sky background, with nine Gaussian components with widths logarithmically spaced from 1 pixel to $\sim$10$\times$ the Kron radius, as recommended by \cite{miller}. We use least-squares minimization to generate best-fit parameters and examine the fits by eye before proceeding with posterior estimation. The posterior distribution and priors are sampled using the Dynamic nested sampling method in \tt dynesty \normalfont initialized from the least-squares minimization \citep{Speagle_2020}. Finally, we calculate the morphological quantities of effective radius, axis ratio, and position angle using a basic analysis function included in \tt imcascade\normalfont.  

As an additional check on our sizes, we model our sample using another popular fitting code, \tt GALFIT \normalfont \citep{Peng_2002}, which utilizes S\'{e}rsic light profiles, following \cite{cutler22}. To mitigate the well-known shortcomings of the uncertainties delivered by size-fitting packages \citep{Haussler_2007}, we add in quadrature a second term in the uncertainty that scales with signal-to-noise ratio (SNR) of the candidates in F160W imaging, as performed previously by \cite{Cutler_2023} and \cite{vdW_12}. The comparison between our \tt imcascade \normalfont and \tt GALFIT \normalfont sizes is shown in Figure \ref{fig:galim}. We find that 90\% of our sample fall within 3$\sigma$ errors of the 1:1 ratio, with the exception of candidate 47006, which returns exceptionally small errors from both \tt imcascade \normalfont and \tt GALFIT\normalfont. In general, we find broad agreement between the two independent size measurement methods, which further supports our assertion that these quiescent candidates are extremely compact. Given this result, we proceed herein with \tt imcascade \normalfont sizes. 

\begin{figure}
\centering
\includegraphics[width=0.5\textwidth]{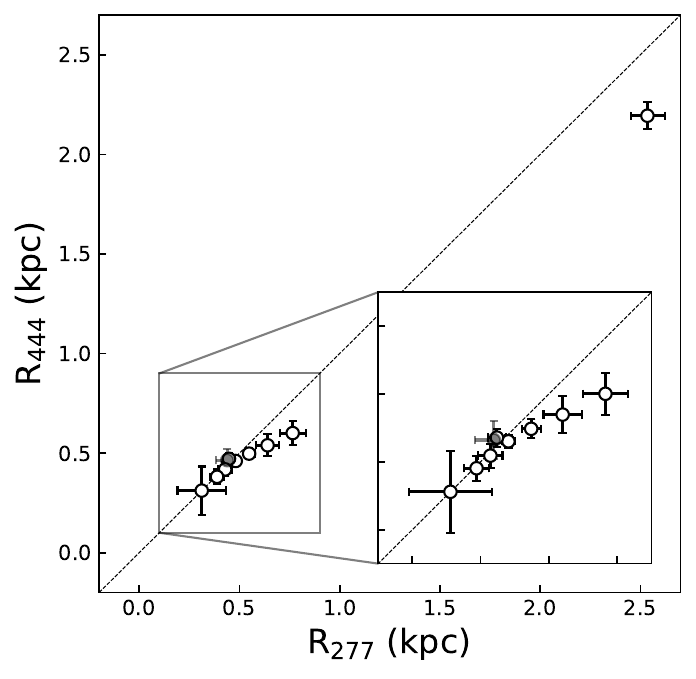}
\caption{F444W vs. F277W \tt imcascade\normalfont-measured sizes for our sample. Seven out of ten (70\%) candidates in the sample exhibit F444W sizes smaller than their F277W sizes, two candidates (20\%) exhibit larger F444W sizes, and one candidate (a point source that we believe to be a possible AGN) does not change in size.} 
\label{fig:svs}
\end{figure}
\section{Results} \label{sec:res}
 
\begin{figure*}
\centering
\includegraphics[width=0.49\textwidth]{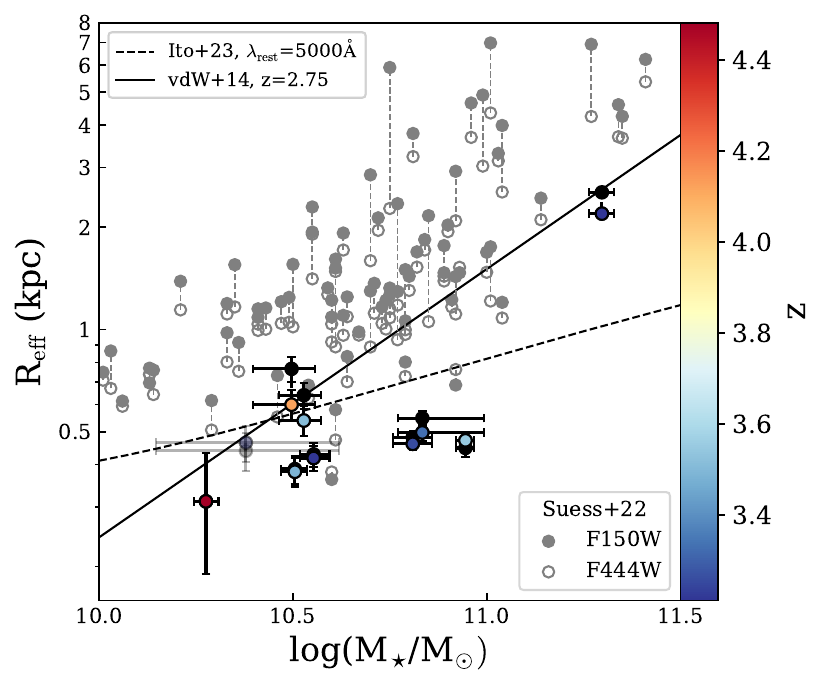}\hfill 
\includegraphics[width=0.51\textwidth]{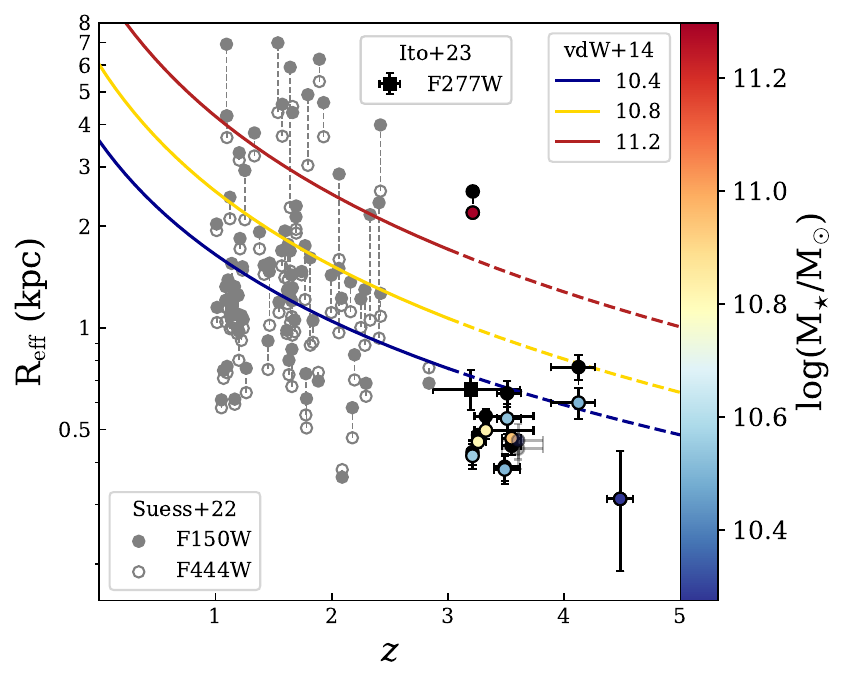}\hfill 
\includegraphics[width=0.54\textwidth]{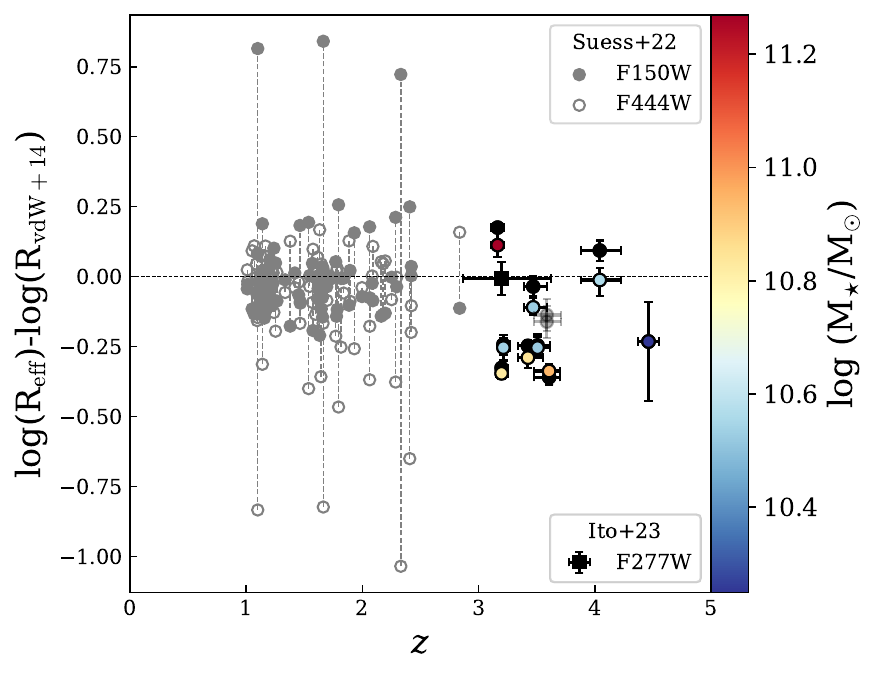}\hfill 
\caption{The key takeaway of this letter is that the measured sizes of eight out of ten (80\%) candidates in our sample are $\sim$40\% smaller than the mass-dependent extrapolation of the rest-optical effective radii of quiescent galaxies out to $z=5$. (Left) We show that the measured sizes of seven out of ten (70\%) candidates in our sample are smaller than both the size-mass relation at z=2.75 defined in \cite{wel} and the best-fit size-mass relation for the UVJ-selected JWST sample defined in \cite{ito23}. The black and colored points represent F277W and F444W sizes, respectively. (Right) From the size-mass distribution of our sample, when extrapolating the size-redshift mass-dependent relative evolution from \cite{wel} to $z=2.5$ (dashed) to the size-redshift distribution of our sample, both color-coded by corresponding stellar mass, the offset remains pronounced. The grey solid and open points represent the F150W and F444W sizes of the quiescent sample from \cite{suess22}, probing similar rest-frame wavelengths (F150W sizes are not significantly offset from earlier relations). (Middle) The difference between the measured size and predicted size in log space when removing the average size-mass relation demonstrates a clear offset towards smaller sizes at $z>3$, not present at lower redshifts or in the ${\lambda}_\mathrm{rest}=5000\mathrm{\AA}$ size of the UVJ-selected quiescent sample from \cite{ito23} (square).}
\label{fig:se}
\end{figure*}

To interpret our results, we first plot the size-mass relation of our sample in the leftmost panel of Figure \ref{fig:se}. The black and colored points represent the F277W and F444W sizes of our sample, respectively. As a baseline, we plot the size-mass relation at $z=2.75$ defined in \cite{wel} (Equation \ref{eqn:sm}) as a solid line:
\begin{equation}
    \mathrm{R_{eff}/kpc=10^{-0.06}(M_{\star}/5{\times}10^{10}M{\odot})^{0.79}}
\label{eqn:sm}
\end{equation}

We additionally plot the best-fit size-mass relation from a recent study that examines the size evolution of quiescent galaxies beyond $z>3$ \citep{ito23} as a dotted line for reference. We observe that seven out of ten (70\%) candidates in our sample exhibit measured sizes smaller than the size-mass relations of both \cite{wel} and \cite{ito23}. We would like to note that while we use Prospector-$\beta$ for our stellar mass estimates, \cite{wel} use \tt FAST \normalfont \citep{FAST} and \cite{ito23} use \tt eazy-py \normalfont \citep{brammer08}, which could, in principle, contribute to some discrepancies in the size-mass relation. However, considering the high median redshift and mass range of our sample, the difference between Prospector-$\beta$-produced and \tt FAST \normalfont-produced stellar mass estimates is only $\sim$0.0$\pm$0.2 dex \citep[Refer to Figure 3 of][]{Leja_2019}.

For further analysis, we compare our results to the measured redshift evolution of the 5000$\mathrm{\AA}$ rest-frame galaxy size for quiescent galaxies at lower redshifts presented in \citet{wel} as a baseline. We extrapolate the size evolution at fixed mass out to $z=5$ to explore how our high redshift sample compares with the extrapolation. We plot the size-redshift distribution of our sample relative to the ${\lambda}_\mathrm{rest}$=5000$\mathrm{\AA}$ size relation from \citet{wel} (Equation \ref{eqn:sz}) for quiescent galaxies (solid lines) in the rightmost panel of Figure \ref{fig:se}
\begin{equation}
    \mathrm{R_{eff}/kpc=B_\emph{z}(1+\emph{z})^{{\beta}_\emph{z}}},
\label{eqn:sz}
\end{equation}

\noindent The extended dashed lines in the rightmost panel of Figure \ref{fig:se} represent the extrapolation of this relation out to $z=5$ for three separate mass bins: log(M$_{\star}$/M$_{\odot}$) = 10.4 (blue), 10.8 (yellow), and 11.2 (red). Coefficients B$_{z}$ and ${\beta}_{z}$ are mass-dependent and can be found in Table 2 of \cite{wel}. The central panel of Figure \ref{fig:se} illustrates the difference between the measured radius and predicted radius in log space when removing stellar mass dependence; points below zero indicate that the measured size is smaller than predicted, and vice versa. We additionally plot the F150W and F444W sizes of the quiescent sample from \cite{suess22} in solid and open grey points, respectively, to comprehensively examine the size evolution of quiescent galaxies over the range of $1<z<5$ based on JWST imaging alone. Eight out of ten (80\%) candidates in our sample exhibit rest-frame optical sizes (F277W) smaller than those extrapolated from Equation \ref{eqn:sm}, with the exception of candidates 89019 and 68124. Additionally, seven out of ten (70\%) candidates in the sample exhibit rest-frame NIR (F444W) sizes smaller than their rest-frame optical sizes (see Figure \ref{fig:svs}), with the exception of candidates 67843 and 71124 (65713 does not change). Of the two candidates with larger F444W sizes (67843 and 71124), only one candidate (71124) appears larger in F444W when measured with both \texttt{GALFIT} and \texttt{imcascade} (see Figure \ref{fig:galim}). Candidate 67843 appears larger in F444W when measured with imcascade, but does not appear larger when measured with \tt GALFIT\normalfont. This candidate also exhibits signs of possible residual or ongoing star formation, which may explain this discrepancy. We note that object 65713, which does not change in size between filters, appears to be unresolved.  Such point sources may be indicative of the presence of Active Galactic Nuclei (AGN), which have been recently uncovered at similar redshifts through JWST  \citep{labbe2023uncover, greene2023uncover, kocevski2023hidden}. 

We find that three of the most massive galaxies in the sample (71124, 82632, 47006) at $\mathrm{log(M_{\star}/M_{\odot})}\sim10.74-10.95$ are surprisingly compact, with rest-frame optical sizes of ${\sim}$0.7 kpc. The average rest-frame optical size for this mass and redshift range extrapolated from \cite{wel}, and confirmed by \cite{ito23}, is $\sim$1.04-1.33 kpc. These sizes are therefore $\sim$0.3 dex smaller than predicted (i.e., $\sim$2$\sigma$ deviations from the size-mass relation). Prior to JWST, high-resolution imaging in rest-frame optical and NIR was not possible, thus we only had access to rest-UV where quiescent galaxies are barely visible. As such, it is only in recent years that we have been able to discover quiescent galaxies at $z>3$ \citep{carnall23, antwidanso2023feniks, Valentino_2023, long23, Gould_2023}. In this letter, we demonstrate that these first quiescent galaxies are overall remarkably compact in both rest-frame optical and NIR imaging. 

\section{Discussion} \label{sec:disc}

In this letter, we measure the effective radii of ten quiescent candidates at $3<z<5$.  We show that eight out of ten (80\%) candidates in the sample exhibit rest-frame optical sizes smaller than those extrapolated from Equation \ref{eqn:sm} from \cite{wel}, with extrapolations extending the original $\sim$6 Gyr baseline by $\sim$1 Gyr or less. This extrapolation agrees well with JWST studies by \cite{ito23}. However, we observe that seven out of ten (70\%) candidates in our sample exhibit measured sizes smaller than the size-mass relations of both \cite{wel} and \cite{ito23}. Moreover, seven out of ten (70\%) candidates in the sample exhibit rest-frame NIR sizes smaller than their corresponding rest-frame optical sizes and one candidate is unresolved in both images. Among these, three of the most massive galaxies in our sample (71124, 82632, 47006) are extremely compact, with rest-frame optical sizes $\sim$~0.7kpc.

Our finding that the majority of our sample appears $\sim$10\% smaller with longer wavelength imaging is consistent with the $\sim$8\% size decrease observed between rest-frame optical (F150W) and rest-frame NIR (F444W) imaging for the $1<z<2.5$ sample of star-forming and quiescent galaxies reported in \cite{suess22}. This size decrease with longer wavelength imaging suggests the presence of negative color gradients in our sample. Negative color gradients could indicate that the central region of the galaxy is dominated by older stellar populations (red), and the outskirts are dominated by younger stellar populations (blue). This is consistent with the `inside out' growth scenario of quiescent galaxies, in which quiescent galaxies slowly gain mass over time via mergers, and therefore accumulate a younger stellar population on the outskirts while retaining an older stellar population in the central region (\citealt{Bezanson09}, \citealt{Naab09}, \citealt{vandeSande13}, \citealt{Hopkins09}). 

However, the central red colors could also be consistent with dust-obscured residual star formation which is commonly observed in star-forming galaxies at $z\sim1-2$ \citep[e.g.,][]{Nelson16}. As similarly massive star-forming galaxies are the natural progenitors \citep[e.g.][]{Toft17}, it stands to reason these negative color gradients may also represent inside-out quenching in at least some cases.  Indeed, spatially resolved studies of quiescent galaxies at slightly lower redshift ($z\sim2$, about 1 billion years later) show a diverse range of formation pathways, with examples of both inside-out and outside-in quenching \citep{Akhshik_2023}.  Evidence for residual dust-obscured star-formation in the cores of these early massive galaxies is confirmed through high-resolution millimeter wavelength detections in some cases \citep{morishita2022compact, lee2023high}, but ruled out in others \citep{Whitaker21}.

The small measured sizes of the first quiescent galaxies herein indicate that the size evolution function of quiescent galaxies does not flatten out and may even be steeper than previously assumed beyond $z=3$. This result is in slight tension with a recently published study by \cite{ito23}, which performs a similar analysis examining the rest-frame 5000$\mathrm{\AA}$ sizes of 26 quiescent candidates first reported by \cite{Valentino_2023}. The conclusion of \cite{ito23} is that quiescent sizes at $z>3$ generally agree with the extrapolation of \cite{wel} out to $z=5$. We illustrate this in Figure \ref{fig:se} by plotting the median redshift and size of the UVJ-selected quiescent sample from \cite{ito23} as a square marker, which clearly shows good agreement with the extrapolation from \cite{wel}. We suspect this discrepancy in our results may be due to the fact that \cite{ito23} use \emph{all} publicly available JWST fields in their study (totaling to $\sim$145 sq. arcmin), while our study focuses specifically on quiescent candidates found in CEERS Epoch 1 Imaging ($\sim$30 sq. arcmin) \citep{Bagley_2023}. The number density of quiescent galaxies in CEERS is predicted to be much higher than the combined JWST fields, indicating that this field may be overdense, which may thus contribute to this different result. \cite{carnall23} report a number density measurement for CEERS of 10.6$^{+4.5}_{-3.4} \times{10^{-5}}$ Mpc$^{-3}$ for $3<z<4$ ($\sim$4.4$\times$ higher than the number density for this redshift range reported in \citealt{Valentino_2023}) and 2.3$^{+3.1}_{-1.5} \times{10^{-5}}$ Mpc$^{-3}$ for $4<z<5$ ($\sim$3.3$\times$ higher than the number density for this redshift range reported in \citealt{Valentino_2023}). Prior to JWST, \cite{peter21} reported a sample of ten spectroscopically-confirmed massive quiescent galaxies at $z\sim3$ in the COSMOS field. The median size of this sample was found to be $\sim2\sigma$ smaller than the extrapolation of the size-redshift function from \cite{wel}. The discovery of this population of compact massive quiescent galaxies at high redshift even prior to JWST imaging further supports our conclusion that the size evolution of quiescent galaxies may be steeper than previously assumed beyond $z>3$. More recently, \cite{carnallnature} reported a spectroscopically-confirmed high-redshift quiescent galaxy at $z=4.658$ that is smaller than previous predictions. This galaxy has an estimated stellar mass of $\mathrm{log(M_{\star}/M_{\odot})}\sim10.58$ and a measured size of $\sim0.2$kpc. This is $\sim$33\% smaller than the \cite{wel} size-mass relation extrapolation of $\sim0.63$ kpc.\\ \citep{peter21}.

Our finding that the majority of rest-NIR sizes are more compact than rest-optical agrees with recent results in the literature \citep{suess22, Miller_2023, vdw23}, with studies reaching a common conclusion that our current understanding of the size growth of quiescent galaxies is biased when only considering $\lambda_{\mathrm{rest}}<1$ $\mu$m. Additionally, our observation that three of the most massive objects in the sample are extremely compact further strengthens the assertion that more massive galaxies are not necessarily larger than their lower-mass counterparts \citep[e.g.,][]{suess22} -- an assumption that has until recently been firmly established in our current understanding of the size-mass relation of quiescent galaxies. Now that we have access to the longer wavelength imaging capabilities of JWST to directly measure the rest-frame NIR sizes of quiescent galaxies up to $z\sim5$, as well as the angular resolution required to spatially resolve their morphologies ($<<$1$^{\prime\prime}$ \citealt{nelson}), our investigation into the size evolution of quiescent galaxies can finally be extended beyond $z>3$. However, this sample has yet to be spectroscopically confirmed. Several upcoming JWST Cycle 2 spectroscopic follow-up campaigns (JWST-GO-2913, JWST-GO-3543, JWST-GO-3567, and JWST-GO-4318) will provide the next step in foremost confirming but also further understanding the nature of these rare high-redshift quiescent galaxies. Until then, this letter aims to  provide a first glimpse into our developing understanding of the size evolution of quiescent galaxies in the distant universe. 

\section{Acknowledgements}
We would like to thank our anonymous referee for providing valuable insight and constructive feedback. This work is based in part on observations made with the NASA/ESA/CSA \emph{James Webb Space Telescope} and the NASA/ESA \emph{Hubble Space Telescope} obtained from the Space Telescope Science Institute, which is operated by the Association of Universities for Research in Astronomy, Inc., under NASA contract NAS 5–26555. The data were obtained from the Mikulski Archive for Space Telescopes at the Space Telescope Science Institute, which is operated by the Association of Universities for Research in Astronomy, Inc., under NASA contract NAS 5-03127 for JWST. These observations are associated with JWST-ERS-1345. These observations are associated with programs HST-GO-12063 and HST-GO-12177. KEW gratefully acknowledges funding from the Alfred P. Sloan Foundation Grant FG-2019-12514. The data products presented herein were retrieved from the Dawn JWST Archive (DJA). DJA is an initiative of the Cosmic Dawn Center, which is funded by the Danish National Research Foundation (DNRF) under grant \#140. Computations for this research were performed on the Pennsylvania State University's Institute for Computational and Data Sciences' Roar supercomputer.

\bibliography{Wright+23.bib}{}
\bibliographystyle{aasjournal}
\end{document}